\documentclass[aps,prl,twocolumn,showpacs]{revtex4}
\newcommand{\hei}[1]{{\bf #1}}
\usepackage{graphicx, amsmath, amsthm}
\makeatletter
\begin{document}

\title{Impurity states in antiferromagnetic Iron Arsenides}

\author{Qiang Han}
\affiliation{Department of Physics,
Renmin University, Beijing, China}

\author{Z. D. Wang}
\affiliation{Department of Physics and
Center of Theoretical and Computational Physics, The University of
Hong Kong, Pokfulam Road, Hong Kong, China}

\date{September 4, 2008}

\begin{abstract}
We explore theoretically impurity states in the antiferromagnetic
spin-density wave state of the iron arsenide. Two types of impurity
models are employed: one has only the intraband scattering while the
other has both the intraband and interband scattering with the equal
strength. Interestingly, the impurity bound state is revealed around
the impurity site in the energy gap for both models. However,  the
impurity state is doubly degenerate with respect to spin for the
first case; while the single impurity state is observed in either
the spin-up or spin-down channel for the second one. The
impurity-induced variations of the local density of states are also
examined.
\end{abstract}


\maketitle

The recent discovery of  iron-based superconductors~\cite{Kamihara}
has triggered intensive efforts to unveil the nature of and
interplay between magnetism and superconductivity in this family of
materials. Series of iron arsenide have been synthesized, which
possess many similar features of the normal and superconducting
states. Experimental measurements have reported that the undoped
ReFeAsO (where Re= rare-earth metals) and AFe$_2$As$_2$ (where
A=divalent metals such as Ba, Ca, Sr) compounds exhibit a long-range
antiferromagnetic spin-density-wave (SDW) order
\cite{Cruz,ChenY,HuangQ,McGuire,Aczel,Goldman}. Upon electron/hole
doping the SDW phase is suppressed and superconductivity emerges
with $T_c$ up to above 50 K
\cite{Takahashi,RenZA,WenHH,ChenGF,ChenXH}.

At present, there is likely certain controversy on the understanding
of the SDW state of the undoped FeAs-based parent compounds. Two
kinds of theories have been put forward: 1) the itinerant
antiferromagnetism, which takes advantage of proper Fermi surface
(FS) nesting (or strong scattering)  between different FS sheets
\cite{DongJ,Mazin,HanQ,Barzykin}; and 2) the frustrated Heisenberg
exchange model of coupled magnetic moments of the localized
$d$-orbital electrons around the Fe atoms
\cite{Yildirim,SiQ,MaFJ,FangC}. As for the itinerant electronic
behavior, first principle band structure calculations \cite{DFT}
based on the density functional theory (DFT) indicate up to five
small Fermi pockets with three hole-like pockets centered around the
$\Gamma$ point and two electron-like ones centered around the $M$
point of the folded Brillouin zone of the FeAs layers, which have
partially supported by the angle-resolved photoemission spectroscopy
(ARPES) from different groups \cite{YangLX,LiuC,ZhaoL,DingH,LuDH}.
Motivated by the DFT calculation and experimental measurements, in
Refs.\cite{HanQ,Barzykin}, the excitonic mechanism \cite{Halperin68}
of itinerant carriers are employed taking account of the FS nesting
between electron and hole pockets and the SDW phase are associated
with triplet excitonic state, which can be understood as condensate
of triplet electron-hole pairs \cite{Halperin68}.

In this paper, we explore theoretically the effect of a single
impurity on the local electronic structure of an Fe-based
antiferromagnet in the triplet excitonic phase. It is shown that
impurity bound states are formed inside the SDW gap, which may be
observed experimentally by local probes. Before introducing the
impurity, we first propose an effective model Hamiltonian to address
the triplet excitonic state,
\begin{eqnarray}
\hat{H}_{MF} &=& \sum_{i,\hei{k},\sigma } \varepsilon_{\Gamma
i}(\hei{k}) d_{i\hei{k}\sigma }^{\dagger} d_{i\hei{k}\sigma}^{}
\nonumber \\
&+& \sum_{\hei{k},\sigma} \varepsilon_X(\hei{k+X})
c_{\hei{k+X}\sigma
}^{\dagger} c_{\hei{k+X}\sigma}^{} \nonumber \\
&+& \sum_{i\hei{k},\sigma,\sigma^\prime}\left[
\Delta_{i\sigma\sigma^\prime}^*(\hei{k})
d_{i\hei{k}\sigma}^{\dagger}c^{}_{\hei{k+X}\sigma^\prime}+H.c.\right],
\label{multiham}
\end{eqnarray}
where $\hei{\Gamma}=(0,0)$, $\hei{X}=(\pi,0)$. We use the index $i$
to label different valence bands around $\Gamma$ point. Around $X$
and $Y$ points, there are two conduction bands. $d_{i\hei{k}\sigma}$
and $c_{\hei{k+X}\sigma}$ are the annihilation operators of
electrons in the $\Gamma i$ and $X$ bands. Theoretically $\hei{X}$
and $\hei{Y}$ are two equivalent nesting directions. Note that, the
structural phase transition occurred just above/on the SDW
transition breaks this equivalency. Without loss of generality, it
is assumed that only conduction band around the $X$ point couples
with the valence bands around the $\Gamma$ point, which is
characterized by the mean-field order parameters
$\Delta_{i\sigma\sigma^\prime}$. For the triplet excitonic phase
(SDW), we have real order parameters satisfying
$\Delta_{i\uparrow\uparrow}=-\Delta_{i\downarrow\downarrow}$ and
$\Delta_{i\uparrow\downarrow}=\Delta_{i\downarrow\uparrow}=0$~\cite{Halperin68}.

\begin{figure}[htb]
\includegraphics[width=8 cm]{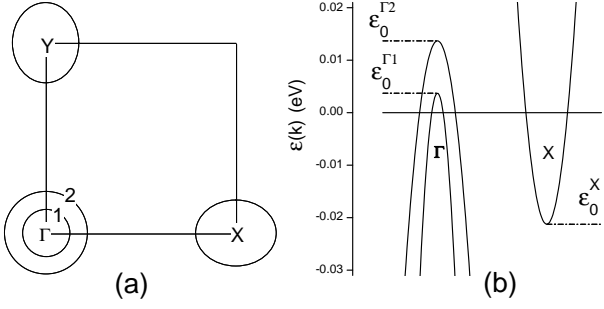}
\caption{Schematic plot of (a) the Fermi surfaces;  and (b) the band
dispersions of the valence (hole) band and conductance (electron)
bands in the unfolded Brillouin Zone for the undoped parent
compound. See text for detail.} \label{fs}
\end{figure}

$\varepsilon_{\Gamma i}(\hei{k})$ and $\varepsilon_X(\hei{k})$ are
used to denote the band dispersions of the nonmagnetic normal state.
For $\hei{k}$ in the vicinity of the $\Gamma$ point (therefore,
$\hei{k+X}$ in the vicinity of the $X$ point), the normal-state
energy dispersions have approximately the 2D parabolic forms
\begin{eqnarray}
&&\varepsilon_{\Gamma i}(\hei{k})= -\frac{\hbar^2(k_x^2+k_y^2)}{2m_{\Gamma i}} + \epsilon_0^{\Gamma i}, \\
&&\varepsilon_X(\hei{k+X})= \frac{\hbar^2(k_x^2+k_y^2)}{2m_X}
-\epsilon_0^X,
\end{eqnarray}
as schematically shown in Fig.~\ref{fs}. Here $m_{\Gamma i}$ and
$m_X$ are the corresponding effective masses. In describing the $X$
band, the elliptic FS is approximated by the circular one for
simplicity. $\epsilon^{\Gamma i}_0$ ($\epsilon^X_0$) denotes the top
(bottom) of the hole (electron) bands. According to the ARPES
measurement\cite{YangLX}, two hole-like Fermi pockets are revealed
around the $\Gamma$ point for undoped BaFe$_2$As$_2$. The band
parameters extracted from the experimental data are as follows.
$m_{\Gamma 1}\approx 2.8 m_e$, $m_{\Gamma 2}\approx 7.4 m_e$, and
$m_X\approx 6.5 m_e$, where $m_e$ is the mass of bare electron.
$\varepsilon_0^{\Gamma 1}\approx 4$ meV, $\varepsilon_0^{\Gamma
2}\approx 16$ meV, and $\varepsilon_0^X\approx 24$ meV. These
parameters indicate that the nesting between the $\Gamma 2$ band and
$X$ band is much better than that of the $\Gamma 1$ band. Therefore
it is natural to assume a larger order parameter $\Delta_{2}$ and a
vanishingly small $\Delta_{1}$. Let $E_g=(\epsilon_0^{\Gamma
2}+\epsilon_0^X)/2$ and $\mu_0=(\epsilon_0^{\Gamma
2}-\epsilon_0^X)/2$. Here $E_G=-2E_g$ denotes the indirect gap
between the top of the $\Gamma$ band and the bottom of the $X$
bands. Therefore, $E_g>0$ describes a semimetal and $E_g<0$ a
semiconductor. With the help of $E_g$ and $\mu_0$ and a further
assumption of $m_{\Gamma 2}=m_X=m \approx 7 m_e$. we can re-express
the energy dispersions as
\begin{eqnarray}
&&\varepsilon_{\Gamma 2}(\hei{k})=-\varepsilon(\hei{k})-\mu_0, \\
&&\varepsilon_X(\hei{k+X})=\varepsilon(\hei{k})-\mu_0, \\
&&\varepsilon(\hei{k})=\frac{\hbar^2}{2m}\hei{k}^2-E_g,
\end{eqnarray}
Note that for $\mu_0=0$, the hole and electron bands are perfectly
nested since $\varepsilon_{\Gamma
2}(\hei{k})=-\varepsilon_X(\hei{k+X})$ and the system is unstable
with respect to infinitesimal Coulomb interaction while for nonzero
$\mu_0$ finite strength of Coulomb repulsion is needed.

For the reason that the order parameter $\Delta_1$ is set to zero,
there is no coupling between the $\Gamma 1$ band and the $X$ band.
The Hamiltonian of Eq.~(\ref{multiham}) is reduced to a model of two
bands with one valence band ($\Gamma 2$ band) and one conduction
band (X band). Introducing the two-component Nambu operator, $
\hat{\psi}_{\hei{k}\sigma}^\dagger=(d_{2\hei{k}\sigma}^\dagger,c_{\hei{k+X}\sigma}^\dagger)
$, the model Hamiltonian can be simplified as
\begin{equation}
\hat{H}_{MF} = \sum_{\hei{k}\sigma}
\hat{\psi}_{\hei{k}\sigma}^\dagger \left(
\begin{array}{cc}
\varepsilon_{\Gamma}(\hei{k}) & \Delta_{\sigma} \\
\Delta_{\sigma} & \varepsilon_{X}(\hei{k})
\end{array}
\right)%
\hat{\psi}_{\hei{k}\sigma} + H_{imp}, \label{twobandHam}
\end{equation}
where an impurity term has been added with the form,
\begin{equation}
\hat{H}_{imp} = \sum_{\hei{k,k^\prime},\sigma}
\hat{\psi}_{\hei{k}\sigma}^\dagger \hat{U}_{\hei{k},\hei{k^\prime}}
\hat{\psi}_{\hei{k^\prime}\sigma},
\end{equation}
where $\hat{U}_{\hei{k},\hei{k}^\prime}$ represents a $2\times 2$
matrix of the scattering potential associated with non-magnetic
impurities. Here we use $\Delta_{\sigma}$ to denote
$\Delta_{\sigma\sigma}$ for short. The Green's function method is
applied to study the single impurity effect. The matrix Greens
functions are defined as
\begin{eqnarray}
\hat{G}_{\sigma\sigma}(\hei{k},\tau;\hei{k^\prime},\tau^\prime) &=&
-\langle T_\tau[\psi_{\hei{k}\sigma}(\tau)
\psi_{\hei{k^\prime}\sigma}^{\dagger}(\tau^\prime] )
\rangle, \\
\hat{G}_{\sigma\sigma}(\hei{k},\hei{k^\prime},i\omega_n) &=&
\int_{0}^{\beta} d\tau
\hat{G}_{\sigma\sigma}(\hei{k},\tau;\hei{k^\prime},0) e^{i\omega_n\tau} \\
\hat{G}_{\sigma\sigma}(\hei{k},\hei{k^\prime},\omega) &=&
\hat{G}_{\sigma\sigma}(\hei{k},\hei{k^\prime},i\omega_n\rightarrow
\omega+i0^+).
\end{eqnarray}
From the Hamiltonian defined in Eq.~(\ref{twobandHam}) we can derive
the bare Green's function
\begin{eqnarray}
\hat{G}_{\sigma\sigma}^0(\hei{k},\omega)&=&\left(
\begin{array}{cc}
\omega-\varepsilon_\Gamma(\hei{k}) & -\Delta_\sigma \\
-\Delta_\sigma  & \omega-\varepsilon_X(\hei{k})
\end{array}
\right)^{-1}, \\
&=& \frac{ \tilde{\omega}\hat{\tau}_0+ \Delta_\sigma
\hat{\tau}_1-\varepsilon(\hei{k})\hat{\tau}_3}{\tilde{\omega}^2-\varepsilon(\hei{k})^2-\Delta_\sigma^2},
\end{eqnarray}
where $\tilde{\omega}=\omega+\mu_0$. $\hat{\tau}_0$ is the $2\times
2$ unit matrix, and $\hat{\tau}_{1,3}$ are the pauli matrices. The
T-matrix approximation is employed to compute the Green's function
in the presence of impurities. For a single impurity, the T-matrix
exactly accounts for the multiple scattering off the impurity. The
single-particle Green's function $\hat{G}$ can be obtained from the
following Dyson's equation,
\begin{eqnarray}
\hat{G}_{\sigma\sigma}(\hei{k},\hei{k^\prime},\omega)&=&
\hat{G}_{\sigma\sigma}^0(\hei{k},\omega)\delta_{\hei{k},\hei{k^\prime}}+
\hat{G}_{\sigma\sigma}^0(\hei{k},\omega) \nonumber \\
&& \hat{T}_{\sigma\sigma}(\hei{k},\hei{k^\prime},\omega)
\hat{G}_{\sigma\sigma}^0(\hei{k^\prime},\omega),
\end{eqnarray}
where the T matrix is given by
\begin{equation}
\hat{T}_{\sigma\sigma}(\hei{k},\hei{k^\prime},\omega)=
\hat{U}_{\hei{k},\hei{k^\prime}}+ \sum_{\hei{k^{\prime\prime}}}
\hat{U}_{\hei{k},\hei{k^{\prime\prime}}}
\hat{G}_{\sigma\sigma}^0(\hei{k^{\prime\prime}},\omega)
\hat{T}_{\sigma\sigma}(\hei{k^{\prime\prime}},\hei{k^\prime},\omega).
\end{equation}
For a point-like scattering potential interacting with itinerant
carriers just on the impurity site, the scattering matrix is
isotropic, $\hat{U}_{\hei{k},\hei{k^{\prime\prime}}}=\hat{U}$. The
above equation is greatly simplified
\begin{equation}
\hat{T}_{\sigma\sigma}(\omega)=\hat{U}+\hat{U}\hat{G}_{\sigma\sigma}^0(\omega)\hat{T}_{\sigma\sigma}(\omega),
\label{tmatrix}
\end{equation}
where
$\hat{G}_{\sigma\sigma}^0(\omega)=\sum_{\hei{k}}\hat{G}_{\sigma\sigma}^0(\hei{k},\omega)
$. After some derivation we obtain
\begin{equation}
\hat{G}_{\sigma\sigma}^0(\omega)= -\pi N_0\left[
\frac{\alpha(\tilde{\omega})}{\sqrt{\Delta_\sigma^2-\tilde{\omega}^2}}(\tilde{\omega}\hat{\tau}_0
+\Delta_\sigma
\hat{\tau}_1)+\gamma(\tilde{\omega})\hat{\tau}_3\right],
\end{equation}
where
\begin{eqnarray}
\alpha(\tilde{\omega})&=&\pi^{-1}\left[\arctan\left(\frac{E_c}{\sqrt{\Delta_\sigma^2-\tilde{\omega}^2}}\right)
\right.+ \nonumber
\\
&&\left. \arctan\left(\frac{E_g}{\sqrt{\Delta_\sigma^2-\tilde{\omega}^2}}\right)\right] \\
\gamma(\tilde{\omega})&=&(2\pi)^{-1}
\ln\left(\frac{E_c^2+\Delta_\sigma^2-\tilde{\omega}^2}{E_g^2+\Delta_\sigma^2-\tilde{\omega}^2}\right),
\end{eqnarray}
with $E_c$ denoting the high-energy cutoff and $N_0=ma^2/(2\pi
\hbar^2)$ the density of states per band per spin. Note that
$\alpha(\tilde{\omega})$ and $\gamma(\tilde{\omega})$ are
independent of the spin index $\sigma$.

The first impurity model we study is the scattering-potential matrix
with only intraband scattering terms, i.e.
$\hat{U}=V_{imp}\hat{\tau}_0$, which was adopted in
Ref.~\cite{Zittartz} to study effect of many impurities. From
Eq.~(\ref{tmatrix}), we obtain
\begin{equation}
\hat{T}_{\sigma\sigma}(\omega)= [\hat{\tau}_0 -
\hat{U}\hat{G}_{\sigma\sigma}^0(\omega)]^{-1}\hat{U} =
[V^{-1}_{imp}-\hat{G}_{\sigma\sigma}^0(\omega)]^{-1}.
\end{equation}
The energy of the impurity bound state is determined by the pole of
$\hat{T}_{\sigma\sigma}(\omega)$, determined by
$\det[V_{imp}^{-1}-\hat{G}_{\sigma\sigma}^0(\Omega)]=0$. Setting
$c\equiv(\pi N_0 V_{imp})^{-1}$, we have the equation for the energy
of impurity bound state
\begin{equation}
c^2+2c\frac{\alpha(\tilde{\Omega})\tilde{\Omega}}{\sqrt{\Delta_\sigma^2-\tilde{\Omega}^2}}
-\alpha(\tilde{\Omega})^2-\gamma(\tilde{\Omega})^2=0. \label{bound1}
\end{equation}
For the spin triplet excitonic phase
$\Delta_{\uparrow}=-\Delta_{\downarrow}$, the above equation gives
rise to impurity states with the same bound energy, i.e. the
impurity states are doubly degenerate. Generally, the above equation
has to be solved numerically to obtain the bound energy
$\tilde{\Omega}$. However, we can get some analytic results under
certain approximations. Under the wide-band approximation $E_c, E_g
\gg |\Delta_\sigma|$, $\alpha(\tilde{\Omega})\approx 1$ and
$\gamma(\tilde{\Omega})\approx \gamma_0=\pi^{-1}\ln(E_c/E_g)$, so we
have
\begin{equation}
\frac{\tilde{\Omega}}{|\Delta_\sigma|}=\mbox{sgn}(c)\frac{1-c^2+\gamma_0^2}{\sqrt{(1-c^2+\gamma_0^2)^2+4c^2}},
\end{equation}
and furthermore if the system has approximately the particle-hole
symmetry $E_c\approx E_g$, then $\gamma_0\approx 0$ and
$\tilde{\Omega}/|\Delta_\sigma|=\mbox{sgn}(c)(1-c^2)/(1+c^2)$ from
the above equation.

For the second impurity model, the  four matrix elements of
$\hat{U}$ is assumed to be the same, i.e. the intra- and inter-band
scattering terms are the same \cite{model2} with
$\hat{U}=V_{imp}(\hat{\tau}_0+\hat{\tau}_1)/2$. Then the T-matrix
according to Eq.~(\ref{tmatrix}) is
\begin{equation}
\hat{T}_{\sigma\sigma}(\omega)=
[2V^{-1}_{imp}-\hat{\tau}^\prime\hat{G}_{\sigma\sigma}^0(\omega)]^{-1}\hat{\tau}^\prime,
\end{equation}
with $\hat{\tau}^\prime=\hat{\tau}_0+\hat{\tau}_1$. The energy of
the impurity bound state is again determined by the pole of
$\hat{T}_{\sigma\sigma}(\omega)$. From $\det[2
V_{imp}^{-1}-\hat{\tau}^\prime\hat{G}_{\sigma\sigma}^0(\Omega)]=0$
we have the equation for $\tilde{\Omega}$,
\begin{equation}
c+\frac{\tilde{\Omega}+\Delta_\sigma}{\sqrt{\Delta_\sigma^2-\tilde{\Omega}^2}}
\alpha(\tilde{\Omega})=0. \label{bound2}
\end{equation}
From the above equation we find that  $\tilde{\Omega}$ is
independent of the function of $\gamma(\tilde{\omega})$ for this
case, which reflects the particle-hole asymmetry. Before solving the
above equation for the bound energy, we study the existence of the
impurity state. Because $\tilde{\Omega}^2<\Delta_\sigma^2$,
$\tilde{\Omega}+\Delta_\sigma$ has the same sign as that of
$\Delta_\sigma$. Therefore, the solution of Eq.~(\ref{bound2})
exists only if the sign of $c$ is opposite to that of
$\Delta_\sigma$. For the SDW state, i.e. the triplet excitonic
phase, we have $\Delta_\uparrow=-\Delta_\downarrow$ and so there is
exactly one impurity bound state in either the spin-up or spin-down
channel. We may  assume
$\Delta_\uparrow=-\Delta_\downarrow=\Delta>0$ as well, then for
attractive scattering $V_{imp}<0$, the impurity bound state only
exists in the spin-up channel and its energy is given by $
\tilde{\Omega}/\Delta=-(1-c^2)/(1+c^2)$ under the wide-band
approximation. If $V_{imp}>0$, however, the impurity state will be
in the spin-down channel, and $
\tilde{\Omega}/\Delta=(1-c^2)/(1+c^2)$. In general, the impurity
bound-state energy is given by
\begin{equation}
\frac{\tilde{\Omega}}{\Delta_\sigma}=\mbox{sgn}(c)\frac{1-c^2}{1+c^2}
\end{equation}
in the valid regime of the wide-band approximation.

To apply the theoretical results to the iron arsenide, we try to pin
down the parameters of our model by extracting them from the
available experimental data for BaFe$_2$As$_2$ \cite{YangLX}.
$\epsilon^\Gamma_0\approx 16$ meV and $\epsilon^X_0\approx24$ meV so
that $E_g\approx 20$ meV. $m_\Gamma\approx m_X\approx 7.0 \ m_e$ and
therefore $N_0\approx 1.2$ eV$^{-1}$.
$\Delta_\uparrow=-\Delta_\downarrow=\Delta\approx 20$ meV. The
high-energy cutoff is set as $E_c=500$ meV, which is of the same
order of magnitude as the band width. Note that $E_g$ extracted from
experimental data is very small, which is in the same order of
magnitude of the order parameter $\Delta$. Therefore, neither the
wide-band approximation nor the particle-hole symmetry can be
applied to the present case. Eqs.~(\ref{bound1}) and (\ref{bound2})
have to be numerically solved.

\begin{figure}[thb]
\includegraphics[width=8 cm]{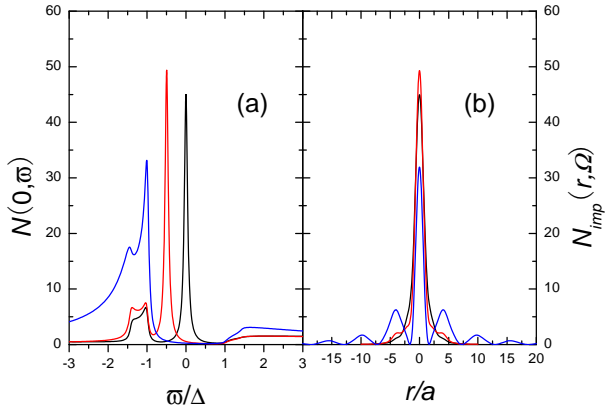}
\caption{LDOS on the impurity site as a function of energy
$\tilde{\omega}$, i.e. $N(0,\tilde{\omega})$ (a), and
impurity-induced LDOS at the bound energy $\tilde{\Omega}$ as a
function of radial distance $r$ off the impurity site, i.e.
$N(r,\tilde{\Omega})$ (b). $r$ is in unit of $a$ with $a$ the
lattice constant of Fe-Fe plane. Black, red, and blue lines
correspond to $V_{imp}=-0.36,-0.6,-4$ eV, respectively.}
\label{ldos-fig}
\end{figure}

Now we examine the local characteristics induced by the impurity by
looking into the variation of the local density of states (LDOS),
which can be probed by the scanning tunneling microscopy (STM). The
LDOS is defined as
\begin{equation}
N(\hei{r},\omega)=-\frac{1}{\pi}\sum_{\sigma}
Im\{Tr[\hat{G}_{\sigma\sigma}(\hei{r},\hei{r}^\prime,\omega)]\},
\label{ldos}
\end{equation}
where $\hat{G}_{\sigma\sigma}(\hei{r},\hei{r}^\prime,\omega)$ the
Green's function in real space. Applying the T-matrix approximation
we have,
\begin{eqnarray}
\hat{G}_{\sigma\sigma}(\hei{r},\hei{r^\prime},\omega)&=&
\hat{G}_{\sigma\sigma}^0(\hei{r},\hei{r^\prime},\omega)+
\hat{G}_{\sigma\sigma}^0(\hei{r},0,\omega) \nonumber \\
&& \hat{T}(\omega)
\hat{G}_{\sigma\sigma}^0(0,\hei{r^\prime},\omega), \label{rspacegrn}
\end{eqnarray}
Substituting Eq.~(\ref{rspacegrn}) into Eq.~(\ref{ldos}) we may
single out the variation of LDOS due to the presence of the impurity
potential,
\begin{eqnarray}
N_{imp}(\hei{r},\omega)&=&-\frac{1}{\pi}\sum_{\sigma}
Im\{Tr[\hat{G}_{\sigma\sigma}^0(\hei{r},0,\omega)
\hat{T}_{\sigma\sigma}(\omega) \nonumber \\
&&\hat{G}_{\sigma\sigma}^0(0,\hei{r},\omega)]\}.
\end{eqnarray}
For the second impurity model\cite{model1}, Fig.~\ref{ldos-fig}(a)
shows the LDOS as a function of energy $\tilde{\omega}$ on the
impurity site, namely $N(0,\tilde{\omega})$, while
Fig.~\ref{ldos-fig}(b) the impurity-induced LDOS at the bound energy
as a function of radial distance $\hei{r}$ off the impurity site,
i.e. $N_{imp}(r,\tilde{\Omega})$. $V_{imp}$ has been set as -0.36,
-0.6, and -4.0 eV, giving rise to the impurity bound states seen as
the sharp peaks  located respectively at the energies
$\tilde{\Omega}/\Delta=0$, $-0.5$, and $-0.99$ in
Fig.~\ref{ldos-fig}(a). The probability densities of these bound
states exhibit a kind of exponential decay with the Friedel
oscillation, as seen in Fig.~\ref{ldos-fig}(b). Introducing two
length scales, $\xi_1$ and $\xi_2$ to characterize the oscillation
and decay, we obtain the asymptotic behavior of
$N_{imp}(r,\tilde{\Omega})$ for large $r$,
\begin{equation}
N_{imp}(r,\tilde{\Omega})\propto r^{-1}\cos^2(\pi
r/\xi_1)\exp(-r/\xi_2).
\end{equation}
$\xi_1/a=\pi/\sqrt{4\pi N_0 E_g}$ which is approximately $5.7$ in
consistence with the numerical results shown in
Fig.\ref{ldos-fig}(b).
$\xi_2/\xi_1=E_g/(\pi\sqrt{\Delta^2-\tilde{\Omega}^2})=0.32$,
$0.37$, and $2.3$ for the three cases of impurity states. This
explains why we see clear Friedel oscillation for impurity state
with bound energy near the gap edge.

\acknowledgments This work was supported by the NSFC grand under
Grants Nos. 10674179 and 10429401, the GRF grant of Hong Kong.


\begin{thebibliography}{20}
\bibitem{model2}
The impurity Hamiltonian is expressed as $H_{imp}=\int
U_{imp}(\hei{r})\phi^\dagger(\hei{r})
\phi(\hei{r})d\hei{r}=\sum_{\hei{k},\hei{k}^\prime}
\hat{\psi}^\dagger(\hei{k})\hat{U}_{\hei{k},\hei{k}^\prime}\hat{\psi}(\hei{k})$
where
$\hat{\psi}^\dagger(\hei{k})=(d^\dagger_\hei{k},c^\dagger_\hei{k+X})$
$\phi(\hei{r})=2^{-1/2}
\sum_{\hei{k}}\varphi_{\Gamma\hei{k}}(\hei{r})d_\hei{k}+\varphi_{X\hei{k}}(\hei{r})c_\hei{k+X}
$, with $\varphi_{\Gamma\hei{k}}(\hei{r})$ and
$\varphi_{X\hei{k}}(\hei{r})$ Bloch functions of quasimomentum
$\hei{k}$. In this paper the scattering potential is assumed to be
pointlike, $U_{imp}=V_{imp}\delta(\hei{r})$. With further
simplification of the Bloch functions of the form
$\exp(i\hei{k}\cdot\hei{r})$, $\hat{U}_{\hei{k},\hei{k}^\prime}$ is
$\hei{k}$ independent and equal to
$V_{imp}(\hat{\tau}_0+\hat{\tau}_1)/2$.
\bibitem{model1}
The following conclusions are also qualitatively correct for the
first impurity model.
\end{thebibliography}
\end{document}